\documentclass[12pt]{article}
\begin{document}
\thispagestyle{empty}
\begin{titlepage}
\thispagestyle{empty}

\begin{flushright}
CERN-TH/98-147\\
 UCLA/98/TEP/13\\
May 1998\\
\end{flushright}

\vskip 2.cm
\begin{center}
{\Large\bf Born-Infeld Corrections to D3 Brane Action in ${\bf AdS_5\times
S_5}$ and N=4, d=4 Primary Superfields.}
\renewcommand{\thefootnote}{\fnsymbol{footnote}}
\footnote{Work supported in part by the EEC under TMR contract
ERBFMRX-CT96-0090, ECC Science Program SCI$^*$ -CI92-0789 (INFN-Frascati), DOE
grant DE-FG03-91ER40662 and CONICIT fellowship}
\vskip 2.cm
{\large S. Ferrara$^{a,b,c}$, M.A. Lled\'o$^{b,d}$ and A. Zaffaroni$^a$}
\end{center}

{\it $^a$ CERN Theoretical Division, CH 1211 Geneva 23, Switzerland.}

{\it $^b$ Physics Department, University of California, Los Angeles. 405
Hilgard Av. Los Angeles, CA 90095-1547. USA.}

{\it $^c$ Institute for Theoretical Physics. Santa B\'arbara,
 CA 93106-4030. USA.}

{\it $^d$  Centro de F\'{\i}sica. Instituto Venezolano de Investigaciones
Cient\'{\i}ficas (IVIC). Apdo 21827 Caracas 1020-A. Venezuela.}

\begin{abstract}

We consider certain supersymmetric Born-Infeld couplings to the D3 brane
action
and show that they give rise to massless and massive KK excitations of
type IIB on $AdS_5\times S_5$, in terms of primary singleton Yang-Mills
superfields.
\end{abstract}
\end{titlepage}

\section{Introduction}

Recently the proposal of an $AdS$/CFT correspondence \cite{malda, pol, witten}, originated by the study of p-brane solitons as interpolating
solutions between maximally symmetric vacua of string and M-theories
\cite{kleb}, has received considerable attention, particularly in relation to
non perturbative aspects of supersymmetric field theories \cite{ooguri}.

In this context, $AdS$ supergravity theories allow to compute correlation
functions of operators of the super Yang-Mills theory, supposedly exact in some
large n limit of the U(n) color gauge group.

This recent analysis is related in an essential way to special properties of
$AdS$ geometries and their boundary at infinity \cite{fr}. Indeed, while
typically the graviton in $AdS_{d+1}$ can be regarded as a unitary irreducible
representation of O(d,2), it can also be represented as a composite boundary
excitation of singleton conformal fields \cite{fer} on the boundary at infinity $\tilde
M_d$, a conformal completion of Minkowski space .

The extension of this correspondence to the interacting singleton theory is the
essence of the proposal of the authors in \cite{malda, pol, witten}
on the $AdS$/CFT correspondence which amounts to compute correlation functions
of conformal composite singleton operators in terms of supergravity Green
functions.

For the case of Type IIB string theory in $AdS_5\times S_5$ \cite{gun2}, related to
D3-brane horizon geometry \cite{malda, pol, witten, kleb}, an infinite sequence of N=4 primary conformal
superfields \cite{FFZ} has recently been identified with the Kaluza-Klein
excitations of type IIB supergravity compactified on $AdS_5\times S_5$
\cite{van, gun}.

Further analysis have shown that the D3-brane coupled to supergravity
background fields enable to correctly reproduce the correlators of marginal
conformal composite operators \cite{fred, seit}.

This analysis has been extended to include also some irrelevant operators which
couple to other background supergravity fields when Born-Infeld corrections
\cite{tsey, tsey2} to the D3-brane action  are taken into account
\cite{gub, das}.

It is worth mentioning that the concept of marginal operator is not invariant
under supersymmetry since in a given supermultiplet relevant, marginal and
irrelevant operators are mixed by supersymmetry.

What is more relevant in the present context is to disentangle some operators
as components of N=4 primary supermultiplets and to look at their
supersymmetric counterpart.

It is the aim of the present paper to extend the analysis of Born-Infeld
corrections to other composite operators and to investigate some supersymmetric
aspects of the non-linear couplings of the D3-brane Born-Infeld action in the
$AdS_5\times S_5$ background.

The paper is organized as follows. In Section 2 we will show how the N=4
primary superfields, corresponding to the short representations of SU(2,2/4),
have a natural interpretation in terms of the Kaluza-Klein spectrum of Type IIB
supergravity on $AdS_5\times S_5$. In Section 3 the N=1 decomposition of N=4
primaries is performed. This allows to analyze Born-Infeld couplings in terms
of N=1 supersymmetry. In Section 4 the Born-Infeld corrections to the D3-brane
action and their supersymmetric properties are studied. An explicit analysis of
all the SU(4) singlet operators occurring in the N=4 primary superfields is
performed
and the analysis for a class of non-singlet operators is outlined. Section 5 ends with some concluding remarks.

\section{ Primary  N=4, d=4  Superfields and Type IIB supergravity in ${\bf
AdS_5\times S_5}$}

The Kaluza-Klein spectrum of Type IIB supergravity compactified on $AdS_5\times
S_5$ was found by Kim, Romans and Van Nieuwenhuizen \cite{van}  using
Kaluza-Klein techniques and by Gunaydin and Marcus \cite{gun} using the
oscillator method to construct unitary irreducible representations of the
SU(2,2/4) superalgebra.

More recently, the same spectrum was analyzed in terms of composite operators
of  N=4 superconformal SU(n) Yang-Mills theory on the boundary and a precise
correspondence between the Kaluza-Klein states and the components of "twisted
chiral" primary superfields was obtained \cite{FFZ}. In this section we will
first  make some remarks on the generic structure of the  order p primary
superfields in relation to the Kaluza-Klein modes coming from higher harmonics
on the five sphere.

Let us first recall that the N=4 primaries \cite{howe} are obtained as suitable
composite operators obtained as polynomials of the singleton superfield
$W_{[AB]}, A B=1,\dots, 4$, satisfying the constraints \cite{ste}
\begin{equation}
W_{[AB]}={1\over 2}\epsilon_{ABCD}\bar W^{[CD]}
\end{equation}
\begin{equation}
\mathcal{D}_{\alpha A}W_{[BC]}=\mathcal{D}_{\alpha [A}W_{BC]}
\end{equation}

An order p primary superfield $A_p=W^p$ is obtained by taking the  trace of the
 order p polynomial of the SU(n) algebra valued singleton superfield projected
on the (0,p,0) SU(4) representation. The $A_p$ superfield corresponds to a
short representation of the SU(2,2/4) superalgebra, including $256\times {1\over
12}p^2(p^2-1)$ states where the factor ${1\over 12}p^2(p^2-1)$ is the dimension
of the (0, p-2,0) SU(4) representation to which the highest spin (two) belongs.
The $A_p$ superfield has the property that its lowest component  is
\begin{equation}
A_p|_{\theta=0}={\hbox{Tr}}(\phi_{\{l_1}\cdots \phi_{l_p\}})-
{\hbox{traces}},\qquad [AB]=l= 1,\dots , 6
\label{sup}\end{equation}
which is the irreducible p-symmetric tensor representation (0, p, 0) of 
SO(6)$\approx$ SU(4). Such polynomial has conformal dimension $E_0=p$. We here
denote the dimension of the conformal operator with the corresponding energy
level $E_0$ of the field representation in $AdS_5$. In the D3-brane
interpretation of the N=4 super Yang-mills theory the scalar field sextet
$\phi_l$ plays the role of the coordinates transverse to the brane
$\phi_l=x^T_l$ (at least in the case where $\phi_l$ belongs to the Cartan
subalgebra of U(N)). By setting $r=\sqrt{{\hbox{Tr}} \phi_l\phi_l}$ we can
define generalized spherical harmonics

\begin{equation}
A_p|_{\theta=0}= r^pY_p(\hat \phi)
\label{primary}\end{equation}
where $Y_p(\hat \phi)$ is in the (0,p,0) representation of SU(4). Indeed
$Y_p(\hat \phi)$ reduces to an ordinary spherical harmonic on $S_5$ in the case
of an U(1) color gauge group.

In this way we find that  any order p primary superfield  contains the
following $D(E_0, J_1,J_2)$ O(2,4) representations

\begin{equation}D(p, 0,0)\qquad\hbox{in}\quad (0,p,0)\label{rep1}\end{equation}

\begin{equation}D(p+1, 1,0)+D(p+1,0,1)\qquad\hbox{in}\quad
(0,p-1,0)\label{rep2}\end{equation}

\begin{equation} D(p+2, 0,0)\qquad\hbox{in}\quad
(0,p-2,0)\label{rep3}\end{equation}

\begin{equation} D(p+2, 1,1)\qquad\hbox{in}\quad
(0,p-2,0)\label{rep4}\end{equation}

\begin{equation}D(p+3, 1,0)+ D(p+3,0,1)\qquad\hbox{in}\quad
(0,p-3,0)\label{rep5}\end{equation}

\begin{equation} D(p+4, 0,0)\qquad\hbox{in}\quad
(0,p-4,0)\label{rep6}\end{equation}

These are the only states in $A_p$ which are in a (0,p,0) SU(4) representation
and which therefore survive when fermions are neglected and only constant
values of the bosonic singleton fields $\phi_l, F_{\mu \nu}$ are retained.

In terms of the singleton fields $\phi_l, F_{\alpha \beta}, F_{\dot \alpha \dot
\beta, }$ ($F_{\alpha \beta }=\sigma_{\alpha \beta}^{\mu \nu}F_{\mu \nu},
F_{\dot \alpha \dot \beta }=\bar F_{\alpha\beta}=\sigma_{\dot \alpha
\dot\beta}^{\mu \nu}F_{\mu\nu}$) the states given by (\ref{rep2}) to
(\ref{rep6}) correspond to the following conformal operators

\begin{equation} \hbox{Tr}(\phi_{\{l_1}\cdots\phi_{l_{p-1}\}}F_{\alpha
\beta})-\hbox{traces}\label{trac1}\end{equation}

\begin{equation} \hbox{Tr}(\phi_{\{l_1}\cdots\phi_{l_{p-2}\}}F_{\alpha
\beta}F^{\alpha \beta})-\hbox{traces}\label{trac2}\end{equation}

\begin{equation} \hbox{Tr}(\phi_{\{l_1}\cdots\phi_{l_{p-2}\}}F_{\alpha
\beta}F_{\dot\alpha \dot\beta})-\hbox{traces}\label{trac3}\end{equation}

\begin{equation} \hbox{Tr}(\phi_{\{l_1}\cdots\phi_{l_{p-3}\}}F_{\alpha
\beta}F^{\alpha \beta}F_{\dot\alpha
\dot\beta})-\hbox{traces}\label{trac4}\end{equation}

\begin{equation} \hbox{Tr}(\phi_{\{l_1}\cdots\phi_{l_{p-4}\}}F_{\alpha
\beta}F^{\alpha \beta}F_{\dot\alpha \dot\beta}F^{\dot\alpha
\dot\beta})-\hbox{traces}\label{trac5}\end{equation}

We observe that in the abelian case (n=1 D3-brane) the previous operators
(\ref{trac1}) to  (\ref{trac5}) would reduce to

\begin{equation}
\matrix{r^{p-1}Y_{p-1}(\hat\phi)F,& r^{p-2}Y_{p-2}(\hat\phi)F^2,&
r^{p-2}Y_{p-2}(\hat\phi)F\bar F, \cr r^{p-3}Y_{p-3}(\hat\phi)F^2\bar F,&
r^{p-4}Y_{p-4}(\hat\phi)F^2\bar F^2& \cr}\label{harm}
\end{equation}

These operators, as we will show in the following, are precisely those who
couple to a particular  spherical harmonic on the $S^5$ sphere corresponding to
a p-wave background field. In this context a state or operator in the (0,p,0)
SO(6) representation will be called a p-wave in connection to its
interpretation with respect to harmonic analysis on a the five-sphere.

{}From the above formulae we observe that a given unitary irreducible
representation of SU(2,2/4), corresponding to a $p$-primary, for $p\geq 4$
mixes p-wave states with $\Delta p=4$ as a consequence of the fact that the
$\theta, \bar \theta$ expansion of the generic twisted chiral superfield  has
components up to $\theta^4 \bar \theta^4$. For $p< 4$ the representation, as
already shown in Ref.\cite{gun} is not generic and in fact we have $\Delta p=3,
\Delta p=2$ for the $p=3$, $p=2$ primaries respectively. From (\ref{trac2}) to
(\ref{trac5}) we can see that  the only primaries which contain SO(6) singlet
states (s-wave) are $p=2$, $p=3$ and $p=4$, giving rise respectively to a spin
2, a complex scalar,  an antisymmetric tensor and a real scalar operator.

{}From the above analysis all non singlet components which occur in the $p>4$
primaries in (\ref{trac2}) to (\ref{trac5}) have the interpretation of
operators which couple to the KK recurrences of the lowest singlet states which
occur for $p\leq 4$.

{}From the explicit analysis of Ref. \cite{van}  the real scalar state of
(\ref{primary}) corresponds to the $p$ wave of the internal components
$a_{\alpha \beta \gamma \delta}$ of the self-dual four form, the real scalar
(\ref{trac5}) corresponds to the $p-4$ wave of the internal part
$h_\alpha^\alpha$ of the metric, the complex scalar (\ref{trac2}) corresponds
to the $p-2$ wave of the dilaton-axion complex $B$ field, the spin 2 state
(\ref{trac3}) corresponds to the $p-2$ wave of the metric $h_{\mu\nu}$ in
$AdS_5$ and finally the antisymmetric tensor states  (\ref{trac1}) and
(\ref{trac4}) correspond  to the $p-1$ and $p-3$ waves of the NS-NS and R-R
antisymmetric tensors $B_{\mu \nu}$.

As shown in Ref.\cite{van} all the states correspond to the same spherical
harmonic $Y^l$ on $S^5$. From the above consideration it then appears explicit
the fact that when the Yang-Mills D3-brane action (or its Born-Infeld
extension) is coupled to the $AdS_5\times S^5$ background the very same term
which originate the coupling  $h^{\mu \nu}T_{\mu \nu}$ will also originate
couplings $h^{\mu \nu}_{(p-2)}T_{\mu \nu}^{(p-2)}$ where $h^{\mu \nu}_{(p-2)}$
is the $p-2$ wave background field and the $T_{\mu \nu}^{(p-2)}$ is the spin 2
operator (\ref{trac3}) contained in the $p$ primary N=4 superfield. The same
considerations apply to the other operators and other background fields  which
are in fact related to the previous one by N=4 supersymmetry.

If we limit ourselves to read the s-wave terms from the Born Infeld action,
they do occur in the $p=2$ primary for $h_{\mu \nu}$ and $B$, in the $p=3$
primary for $B_{\mu\nu}$ and in the $p=4$ primary for $h_\mu^\mu$.
Interestingly enough, N=4 supersymmetry predicts the non linear corrections to
the quadratic Yang Mills actions from the structure of the N=4 primary
superfields. It should be also noted that $B_{\mu \nu}$ also couples linearly
to the Yang-Mills field strength as shown in (\ref{trac1}), but this operator
comes in the l=1 wave as component of the $p=2$ primary supercurrent multiplet \cite{howe, cec, roc, zum2}.

Finally, we remark that the above analysis is complete as far as only constants
$\phi_l, F_{\alpha\beta}$ occur. The generalization to all other states, to
include fermions as well as non constant $\phi_l$ is in principle implemented
by N=4 conformal symmetry. It will be shown in the following sections that the
Born-Infeld couplings of the D3-brane action in the $AdS_5\times S^5$
 background
exactly produce the terms discussed in the body of this section.

\section{N=1 decomposition of N=4 primary superfields}

In order to examine the supersymmetric structure of the Born-Infeld non linear
action, it is useful to decompose the N=4 primary superfields in N=1 components
since the analysis of the four dimensional Born-Infeld action is mostly known
from an N=1 perspective \cite{cec, des, roc}.

{}From a group theoretical point of view, the N=1 decomposition of N=4
superfields amounts to decompose the SU(2,2/4) superalgebra and its
representations under the U(2,2/1)$\times$SU(3)$\times$U(1). The decomposition
of the N=4 Yang-Mills superfield strength in N=1 parts is \cite{zum}
\begin{equation}
W_{[AB]}=(S_i,W_\alpha)
\end{equation}
where $S_i$ is a triplet of chiral N=1 superfields and $W_\alpha$ is the chiral
N=1 field strength. These superfields have U(1) charge 1 and ${3\over 2}$
respectively. All these superfields are supposed to be U(n) Lie algebra valued.
It is now straightforward to obtain the N=1 decomposition of the N=4 twisted
chiral superfields. It suffices to decompose the (0,p,0) SU(4) representation
into SU(3) representations, using the fact that $6\mapsto 3+\bar 3$ so that the
real scalar sextet $\phi_l$ goes into $S_i,\bar S_i$ while $W_{\alpha}$ is
SU(3)  inert.

Let us first consider the $p=2$ primary corresponding to the graviton multiplet
in $AdS_5$. Its N=1 superfield content is

\begin{equation}
S_iS_j,\quad \bar S_i\bar S_j, \quad S_i\bar S^j- {1\over 3}\delta^j_iS_kS^k
\label{decomp1}\end{equation}
\begin{equation}
S_iW_\alpha,\quad \bar S_i\bar W_{\dot\alpha}
\label{decomp2}\end{equation}
\begin{equation}
S_i\bar W_{\dot \alpha},\quad \bar S_i\ W_{\alpha}
\label{decomp3}\end{equation}
\begin{equation}
W_\alpha^2,\quad \bar W_{\dot\alpha}^2
\label{decomp4}\end{equation}
\begin{equation}
W_{\alpha}\bar W_{\dot \alpha}
\label{decomp5}\end{equation}

Note that the 20 scalars of (\ref{primary}) are in (\ref{decomp1}) since under
SU(4)$\mapsto$SU(3) $20\mapsto 6+\bar 6 + 8$. The fifteen SU(4) currents are in
(\ref{decomp1}), (\ref{decomp3}), (\ref{decomp5}) since $15\mapsto 8+3+\bar 3 +
1$.
The 8 spin 3/2 supercurrents are in (\ref{decomp3}),(\ref{decomp5}) since
$4+\bar 4\mapsto 3+\bar 3 + 1+\bar 1$. The 12 antisymmetric tensors are in
(\ref{decomp2}),(\ref{decomp3}) since $6_c\mapsto (3+\bar 3)_c$ and finally the
stress tensor singlet is in (\ref{decomp5}).

The above results apply for constant values of the $S_i|_{\theta =0}$ component
and vanishing fermions. The stress tensor in (\ref{decomp5}) would otherwise
receive an extra contribution.

At this point we notice  that the conformal supergravity multiplet  is
contragradient  to the supercurrent multiplet \cite{ste, FFZ, zum2}. Then we learn that its SU(4) singlet part  reduces to the N=1
superconformal gravity potential  $H_{\alpha\dot\alpha}$ \cite{zum2} and to an
extra chiral superfield $S$.

We now move to the $p=3$ primary superfield. The 50 scalars in the (0,3,0)
SU(4) representation correspond to
\begin{equation}
S_iS_jS_k,\quad S_i\bar S_j\bar S_k-\hbox{traces}, \bar
S_iS_jS_k-\hbox{traces},\quad \bar S_i\bar S_j\bar S_k
\end{equation}
This correspond to the SU(3) decomposition $50\mapsto 10 +\bar 10+15 +\bar
{15}$. The other N=1 component superfields are

\begin{equation}
S_iW_\alpha^2,\quad \bar S_iW_\alpha^2,\quad \bar S_i\bar W_{\dot
\alpha}^2,\quad  S_i\bar W_{\dot \alpha}^2
\label{axion}
\end{equation}
\begin{equation}
S_iW_\alpha\bar W_{\dot \alpha},\bar S_iW_\alpha\bar W_{\dot \alpha}
\end{equation}
\begin{equation}
W_\alpha^2\bar W_{\dot \alpha},\quad \bar W_{\dot \alpha}^2W_\alpha
\end{equation}
\begin{equation}
S_iS_jW_\alpha,\quad (S_i\bar S^j-{1\over 3}\delta_i^j\bar S_k\bar
S^k)W_\alpha, \quad
\bar S_i\bar S_jW_\alpha
\end{equation}
\begin{equation}
S_iS_j\bar W_{\dot \alpha},\quad (S_i\bar S^j-{1\over 3}\delta_i^j\bar S_k\bar
S^k)\bar W_{\dot \alpha}, \quad
\bar S_i\bar S_j\bar W_{\dot \alpha}
\end{equation}
We can proceed further. For the $p=4$ primary superfield the (0,4,0) SU(4)
representation decomposes under SU(3) as follows
\begin{equation}
105\mapsto 15+ \bar{15} +24+ \bar{24}+27
\end{equation}
This corresponds to the N=1 polynomials
\begin{equation}
\matrix{S_iS_jS_kS_l,&S_iS_jS_k\bar S_l-\hbox{traces},&S_iS_j\bar S_k\bar
S_l-\hbox{traces},\cr
S_i\bar S_j\bar S_k\bar S_l-\hbox{traces},&\bar S_i\bar S_j\bar S_k\bar S_l}
\end{equation}
Superfield multiplication with the $W_\alpha$ gives the rest of the superfields
as before. For instance, the stress tensor recurrence sits in the N=1
superfields
\begin{equation}
S_iS_jW_\alpha\bar W_{\dot \alpha},\quad \bar S_i\bar S_jW_\alpha\bar W_{\dot
\alpha}, (S_i\bar S^j-{1\over 3}\delta_i^jS_k\bar S^k)W_\alpha\bar W_{\dot
\alpha}
\label{graviton}\end{equation}
Note that the  new superfield $W_\alpha^2\bar W_{\dot \alpha}^2$ appears.

 The above superfields give the supersymmetric completion of the component
fields given by (\ref{trac1}) to (\ref{trac5}). We see that $p$ waves
superfields correspond to a dependence on the $S_i$ multiplet. If we want to
confine only to s wave superfields we can se $S_i=0$ and then only $p=2,3,4$
primaries remain with N=1 components given by
\begin{equation}
W_\alpha^2,W_\alpha\bar W_{\dot \alpha},W_\alpha^2\bar W_{\dot
\alpha},W_\alpha^2\bar W_{\dot \alpha}^2\label{swave}
\end{equation}
The bosonic components of these singlets operators are

\begin{equation}
W_\alpha^2\mapsto O_2={1\over 2}(F^2\pm F\tilde F)\qquad (\tilde F_{\mu
\nu}={i\over 2}\epsilon_{\mu \nu \rho \sigma}F^{\rho \sigma})
\label{qui1}\end{equation}
\begin{equation}
W_\alpha\bar W_{\dot \alpha}\mapsto T_{\mu \nu}=F_{\mu \nu}F_{\nu \rho}-
{1\over 4}\eta_{\mu \nu}(F_{\sigma \rho})^2
\label{qui2}\end{equation}
\begin{equation}
W_\alpha^2\bar W_{\dot \alpha}\mapsto O_3=T_{\mu \rho}F_{\rho \nu}=F_{\mu
\sigma}F_{\rho \sigma}F_{\rho \nu}- {1\over 4}(F_{\sigma \rho})^2F_{\mu \nu}
\label{qui3}\end{equation}
\begin{equation}
W_\alpha^2\bar W_{\dot \alpha}^2\mapsto O_4={1\over 4}[(F^2)^2-(F\tilde F)^2]=
F_{\mu \rho}F_{\nu \rho}F_{\mu \sigma}F_{\nu \sigma}- {1\over 4}(F^2)^2
\label{qui4}\end{equation}

The above expression strictly apply for a U(1) gauge theory. Suitable
symmetrizations must been understood when a non abelian trace is taken \cite{tsey2}.

All other higher order primaries are accordingly obtained by suitable chiral and antichiral multiplication of the superfields $S_i, W_\alpha$.

Note that these multiplications only generate field components with spin $\leq 2$, as follows from a general N=1 superfield with at most one external vector index.

In particular we note that an N=4 twisted chiral superfield is not chiral when reduced to N=1 or N=2 supersymmetry. Therefore the KK towers identified in \cite{witten} only reproduce part of the entire KK spectrum on $AdS_5\times S_5$.

\section{The ${\bf AdS}$/CFT correspondence}
The relation between $N=4$ SYM and type IIB on $AdS_5\times S^5$
predicts that there is a one-to-one correspondence between the supergravity KK
modes and the SYM composite operators belonging to  short multiplets. The
$N=4$ covariant description of this correspondence was discussed  in
\cite{FFZ}. Given the mass and the $SU(4)$ quantum number of the supergravity
mode, the
corresponding SYM operator can be uniquely identified as a component
of one of the superfields $A_p=\hbox{Tr }W^p$ as defined in (\ref{sup}).

It has been proposed in \cite{das} that the SYM operator corresponding to
a given supergravity mode can be also be identified by expanding the
Born-Infeld lagrangian around the anti-de-Sitter background.  We now give evidence that  there is a complete agreement between the
two methods.

The Born-Infeld lagrangian can be generalized to include the couplings to
the type IIB supergravity fields. Considering only the bosonic terms, the
coupling to the NS-NS fields ($\phi,B^{\hbox{NS}}$) can be written as
\begin{equation}
L_{\hbox{BI}}= \sqrt{-{\rm det}\left (g_{\mu\nu}+ e^{-\phi /2}{\cal
F}_{\mu\nu}\right)}
\label{lbi}\end{equation}
where ${\cal F}=F-B^{\hbox{NS}}$ \cite{witD}, while the coupling to the RR fields
$(\tilde\phi, B^{\hbox{RR}}, A^{(4)})$ requires the introduction of a Wess-Zumino term
\cite{douglas},
\begin{equation}L_{\hbox{WZ}} = A^{(4)} + B^{\hbox{RR}}\wedge {\cal F} + \tilde\phi {\cal
F}\wedge {\cal F}
\label{WZ}\end{equation}

We will focus on the states discussed in Section 2, which can be constructed
in terms of the $N=4$ gauge fields and constant scalars.
The generalization of the composite operators, which will be identified in this
section, to include the fermions and derivatives of the scalars is dictated by
the $N=4$ supersymmetry.

We first discuss the three $SU(4)$ singlets (besides the graviton) contained in
the KK tower and given by (\ref{qui1}),(\ref{qui3}) and (\ref{qui4}). The
corresponding bosonic operators are a complex scalar, an antisymmetric
tensor and a real scalar.

The dimension $E_0$ of a scalar SYM operator is related to the mass of the
corresponding supergravity mode by the formula \cite{fr,fer}
\begin{equation}
m^2=E_0(E_0-4)
\label{dimension}
\end{equation}
while for an antisymmetric tensor we have \cite{FFZ}
\begin{equation}
m^2=(E_0-2)^2
\label{dimanti}
\end{equation}

$A_2=\hbox{Tr}W^2$ gives the two scalars in (\ref{qui1}) with dimension
$E_0=4$. They couple respectively, to the type IIB
dilaton via the Born-Infeld lagrangian, and the RR scalar via the Wess-Zumino
term. The corresponding N=1 background superfield  was called $S$ in the
previous section. The full multiplet of currents corresponds indeed to the
graviton
multiplet in $AdS_5$ \cite{fer,FFZ}, which contains a massless complex singlet
scalar (the dilaton and the RR scalar), in agreement with eq.
(\ref{dimension}).

The explicit form of the singlet operators in $A_3=\hbox{Tr}W^3$ and
$A_4=\hbox{Tr}W^4$ is given by (\ref{swave}) (using an $N=1$ superfield
formalism). Their bosonic components, as given by (\ref{qui3}) and
(\ref{qui4}), have dimension $E_0=6,8$ respectively.

Let us start with the latter. Using formula (\ref{dimension}), we see that it
corresponds to a scalar supergravity mode with squared mass 32.
The full KK spectrum of type IIB on $AdS_5\times S^5$ was computed in
\cite{van} and a singlet scalar with squared mass 32 was identified (see fig. 2
and
table III in \cite{van}). The corresponding type IIB field is the dilatational
mode of the internal ($S^5$) metric $h^\alpha_\alpha$. In the higher harmonic
modes in this KK tower, $h^\alpha_\alpha$ mixes with the scalar obtained by
reducing the type IIB four-form along $S^5$. However, the lower state, which
corresponds to the singlet with squared mass 32, is not coupled to the
four-form. The only other non-zero type IIB field is the Weyl mode of the
space-time metric
$g_{\mu\mu}$. The coupled equations are (eq. (2.21) and (2.40) in \cite{van})
\begin{equation}
(\partial^2 - 32)h^\alpha_\alpha =0\,\,\,\,\,\,\,\,\,\,\,\,\,
g_{\mu\mu}-{8\over 3} h^\alpha_\alpha =0
\label{equation}
\end{equation}
According to the proposal in \cite{das}, we should now expand the Born-Infeld
lagrangian in the $AdS$ background, perturbed with  non-trivial
$g_{\mu\mu},h^\alpha_\alpha$, satisfying eq. (\ref{equation}). Let us call
$\pi=g_{\mu\mu}=(8/3)h^\alpha_\alpha$. $h^\alpha_\alpha$ enters in the
Born-Infeld action when we
compute the pullback of the type IIB metric; it is coupled to terms with SYM
scalar derivatives and fermions that we put, for simplicity, to zero. $\pi=
g_{\mu\mu}$ is
the only field that couples to $F_{\mu\nu}$. It is the first component of a
chiral superfield which we denote by $H$.

The expansion of $L_{BI}$ in (\ref{lbi}) up to quartic order in the super
Yang-Mills operators and in flat background reads

\begin{equation}
L_{BI} = {1\over 2}F^2 -{1\over 8} O_4 + \cdots
\label{born}
\end{equation}
where the higher dimensional operators have been supressed. The term $F^2$
is clearly conformal invariant. Therefore the leading operator that couples
to the Weyl mode $\pi$ is exactly the operator $O_4$, in agreement with the
expectations. The expansion has been done, strictly speaking, for an abelian
theory. Here and in the following, a suitable symmetrization in color space
must be understood in the non-abelian case.

 The N=1 supersymmetric generalization of (\ref{born}) in N=1 superbackground
with $S, H_{\alpha\dot\alpha}, H$ turned on is
\begin{equation}
SW_\alpha^2|_F+H^{\alpha\dot\alpha}W_\alpha\bar
W_{\dot\alpha}|_D+HW_\alpha^2\bar W_{\dot\alpha}^2|_D+\cdots
\label{bornin}
\end{equation}
Note that the first term of the above expression also contains the last term of
(\ref{WZ}).

Let us now consider the operator $O_3$ with dimension 6. It should correspond
(using equation (\ref{dimanti})) to a singlet antisymmetric tensor in
$AdS_5$ with squared mass 16. It can be found in fig.3 and table III in
\cite{van}; it comes from a combination of the modes of the NS-NS and R-R type
IIB antisymmetric
tensors which are constant along the five-sphere.
The expansion of the Born-Infeld lagrangian in the $AdS_5$ background perturbed
by non-zero type IIB antisymmetric tensors was performed in \cite{raj,das},
where only part of the operator $O_3$ was found. That the full expression in
(\ref{qui3}) must appear is predicted by supersymmetry and it is
easily derived from the lagrangian (\ref{born}). Ignoring SYM scalars and
fermions, the only modification due to the non-trivial background amounts to
substitute $F$ with $F-B$ in eq. (\ref{born}). Here B is the NS-NS type IIB
antisymmetric tensor. The R-R tensor is determined by the equation of motion to
be
$B^{\hbox{RR}}_{\mu\nu}={1\over 2} \epsilon_{\mu\nu\tau\rho}B^{\tau\rho}$
\cite{van,das}. The RR tensor appears in the D3 action via the Wess-Zumino
term, and gives
couplings of the form $F_{\mu\nu}B^{\mu\nu}$ which are subleading in the large
n limit.
The Wess-Zumino term is however crucial, as notice in \cite{das}, in cancelling
dimension four operators, constructed with fermions, which couple to B in the
Born-Infeld lagrangian.

Substituting F with F-B in the Born-Infeld lagrangian (\ref{born}), expanding
at the first order in B and neglecting terms which are subleading for large N,
we explicitely find the coupling $BO_3$.

The $N=1$ supersymmetric generalization of the operator $O_3$ is easily
obtained by promoting $B_{\mu\nu}$ to a chiral superfield  $L_\alpha$
\cite{fzw} and by shifting $W_\alpha\mapsto W_\alpha-L_\alpha$ in
(\ref{bornin}). From  the quartic term in the
Born-Infeld lagrangian we then obtain the coupling

\begin{equation}
\bar W_{\dot\alpha}^2 W_\alpha L^\alpha + {\rm h.c}
\end{equation}

 It should be noted that, as observed in \cite{das}, it is crucial to expand
the lagrangian in the $AdS_5$ background. This means that the metric along
the D3 brane is flat, but depends on the fifth coordinate r of $AdS$ as
$g_{\mu\nu}={r\over R}\eta_{\mu\nu}$, where R is the $AdS$ radius. The D3 brane
is assumed to live at a fixed value of r. In the expansion around this
background,
all the coupling between supergravity fields and SYM operators are dressed by
powers of r/R.

Having identified the three SU(4) singlets in the Born-Infeld expansion, we
can turn to the higher harmonics. Consider first the abelian case. The scalar
field sextet $\phi_l$ describe the
position of the D3 brane in the transverse space. Our prescription will be to
expand the action
of a single D3 brane in the $AdS_5$ background, keeping the position r fixed
and indentifying the coordinates on $S^5$ (which appear in a p-wave operator as
the spherical function $Y_p$) with the scalar fields $\hat \phi_l=\phi_l/r$.
With this identification, the harmonic of degree p in the KK expansion
of the dilaton, for example,  gives rise to a coupling to the
operator $O_2\times Y_p(\hat\phi)$, where $Y_p$ is the spherical harmonic of
degree p.

Let us now outline the identification of all the other states in (0,p,0)
representations of SU(4). According to the analysis in Ref. \cite{van}, the
real scalar state of (\ref{primary}) corresponds to the $p$ wave of the
internal components $a_{\alpha \beta \gamma \delta}$ of the self-dual four
form. It can indeed be obtained using the first term in eq. (\ref{WZ}),
because, as noticed in Ref. \cite{van}, the components of the four-form along
the D3 brane are determined in terms of the internal ones by the equations of
motion.  The expansion of the Born-Infeld lagrangian in the Weyl mode of the
metric gives the real scalars (\ref{trac5}) corresponding to the higher
harmonics of the internal part $h_\alpha^\alpha$ of the metric. The equations
of motion found in Ref. \cite{van} imply indeed the
relation $g_{\mu\nu}={16\over 15}h^\alpha_\alpha$.
Finally, the two antisymmetric tensor states  (\ref{trac1}) and (\ref{trac4}),
corresponding  to modes of the NS-NS and R-R  antisymmetric tensors, can be
obtained by expanding both the Born-Infeld lagrangian and the Wess-Zumino term
around the solution of the equation of motion for $B^{\hbox{NS}}$ and $B^{\hbox{R}}$. We obtain
contributions of the form,
\begin{equation}
L_{\hbox{BI}}= B^{\hbox{NS}}Y_p(\hat\phi)(F+O_3),\,\,\,\,\,\,\,\,
L_{\hbox{WZ}}= B^{\hbox{RR}}Y_p(\hat\phi)\wedge F
\label{two}\end{equation}
Notice that the terms $BFY_p$ are no more subleading in the large n limit.
According to the equations of motion discussed in Ref.\cite{van}, the p-wave
components of the antisymmetric tensors, when $p\ge 1$, give two different
modes, satisfying $B^{\hbox{RR}}=\pm *B^{\hbox{NS}}$, with squared mass
$p^2$ and $(p+4)^2$. This splitting in the masses can be understood by
observing that eq.(\ref{two}) gives $BFY_p$ for one of the modes, while for the
other there is a cancellation between $L_{\hbox{BI}}$ and $L_{\hbox{WZ}}$, and the first
non-trivial coupling occurs for an
higher dimensional operator, $BO_3Y_p\,$\footnote{The same splitting was
noticed in the context of the absorption of supergravity modes by the D3 branes
in \cite{mathur}.}.

Notice that, as discussed in Ref. \cite{van}, the $p=0,1$ modes for
$a_{\alpha\beta\gamma\delta}$ and one of the $p=0$ modes for $B_{\mu\nu}$ are
zero or can be gauged away, in agreement with the fact that $\hbox{Tr}\phi_l$ and
$\hbox{Tr}F_{\mu\nu}$
do not belong to the series of composite operators in $A_p$.

 It is easy to check, in
all the cases discussed above, that the dimensions of the operators $O_iY_p$
correspond, via the eqs.(\ref{dimension}) and (\ref{dimanti}), to the masses
computed in Ref.\cite{van}.

It should be possible to extend the present analysis to include all the
remaining states in the superfields $A_p$ in SU(4) representations other than
(0,p,0). In general, they involves fermions and scalar derivatives. The
structure of the 15 currents of SU(4), associated with the massless gauge
fields in the graviton multiplet in $AdS_5$, for example, is easily seen to
emerge from
the coupling of the D3 brane action to the mixed components of the metric
$g_{\mu\alpha}$ along the $S^5$ Killing vectors. The KK recurrences of the SU(4) adjoint currents are states in the (1, p-2, 1), (1, p-3, 1) and (1, p-4, 1) SU(4) representations. They correspond to the $D(p+1, 1/2, 1/2), D(p+2, 1/2, 1/2)$ and $D(p+3,  1/2, 1/2)$ O(2,4) conformal fields respectively. It is likely that, also
for all these states, the composite operators, which can be determined by
expanding the Born-Infeld action around the corresponding supergravity modes,
will
agree with what is predicted by superconformal invariance, in terms
of components of some superfield $A_p$.

 The previous analysis is in
agreement with the discussion in section 2 in the abelian case. As discussed in
details in section 2, the composite SYM operators, that we expect to be
associated to these supergravity modes in the non-abelian case, are the
non-abelian generalizations of the operators $O_i\times Y_p(\hat\phi)$ which
are contained in the
superfields $A_p$.
It is likely that the color structure will
emerge naturally from a full $N=4$ non-abelian generalization of the
Born-infeld action and from a better understanding of how the coordinates of n
D3 branes
can be promoted to SU(n) matrices \cite{tsey2}.
{}From the present analysis, it should be possible to extend the Born-Infeld
action to include the dependence on the non singlet fields $S_i$. In this case
the KK graviton recurrences should couple to the conformal operators as given
in (\ref{graviton}) while the dilaton-axion recurrences should be given by
operators  like (\ref{axion}). It is clear that these operators must come from
the N=4 completion of (\ref{bornin}).

\section{Conclusions}

In this paper, we expanded the Born-Infeld action around the $AdS_5\times S^5$
background and we found that certain KK supergravity modes exactly couple to
the composite operators predicted by the SU(2,2/4) algebra, and
studied in \cite{FFZ}. It is likely that this correspondence, with a correct
description of how the D3 brane is embedded in the non-trivial $AdS$ geometry
and a better understanding of the non-abelian and $N=4$ supersymmetric
structure of the Born-infeld action, can be extended to the entire KK spectrum.
Information about the composite operators associated to KK modes can be
obtained also by studying the absorption of the supergravity fields by the D3
branes \cite{absor, gub, mathur}. This method makes use of the full
D3 brane metric
and not only of the near-horizon geometry, and it would be interesting to
understand the relation with the approach considered in this paper.

\section{Acknowledgments}

One of us (S. F.) would like to acknowledge  A. Tseytlin for interesting discussions and the 
 Institute of Theoretical Physics in Santa Barbara, where part of this work was done, for its kind hospitality durind the String Duality Institute.

\end{document}